\documentclass[onecolumn]{aa} % for a paper on 1 column  
\usepackage{graphicx}
\usepackage{txfonts}
\begin{document}

   \title{Comets $^{12}$CO$^+$ and $^{13}$CO$^+$ fluorescence models for
   measuring the $^{12}$C/$^{13}$C isotopic ratio in CO$^+$}

    \titlerunning{CO$^+$ fluorescence spectrum}

%   \subtitle{I. Overviewing the $\kappa$-mechanism}

   \author{P. Rousselot\inst{1}
          \and
          E. Jehin\inst{2}
          \and
          D. Hutsem\'ekers\inst{2}
          \and
          C. Opitom\inst{3}
          \and
          J. Manfroid\inst{2}
          \and
          P. Hardy\inst{1,4}
          }

   \institute{Institut UTINAM - UMR 6213, CNRS / Université de Franche-Comt\'e, OSU THETA, 41 bis Av. de l'Observatoire, BP 1615, F-25010 Besançon Cedex, France\\
   \email{philippe.rousselot@obs-besancon.fr}
        \and
        STAR Institute, Univ. of Li\`ege, All\'ee du 6 Ao\^ut 19c, 4000 Li\`ege, Belgium\\
         \and
         Institute for Astronomy, Univ. of Edinburgh, Royal Observatory, Edinburgh EH9 3HJ, UK\\
        \and 
        Laboratoire Interdisciplinaire Carnot de Bourgogne - UMR 6303, CNRS / Universit\'e de Bourgogne, 9 Av. A. Savary, BP 47870, F-21078 Dijon Cedex, France
             }

   \date{Submitted 20 Sept 2023 to Astronomy \& Astrophysics}

% \abstract{}{}{}{}{} 
% 5 {} token are mandatory
 
  \abstract
   {CO is an abundant species in comets, creating CO$^+$ ion with emission lines that can be observed
   in the optical spectral range. A good modeling of its fluorescence spectrum is important
    for a better measurement of the CO$^+$ abundance. Such a species, if abundant enough, can also be used to
   measure the $^{12}$C/$^{13}$C isotopic ratio.}
   {This study uses the opportunity of a high CO content observed in the comet C/2016 R2 
   (PanSTARRS), that created bright CO$^{+}$ emission lines in the optical range, to 
   build and test a new fluorescence model of this species and to measure
   for the first time the $^{12}$C/$^{13}$C isotopic ratio in this chemical species with
   ground-based observations.} 
   {Thanks to laboratory data and theoretical works available in the scientific literature 
   we developed a new fluorescence model 
  both for $^{12}$CO$^+$ and $^{13}$CO$^+$ ions. The $^{13}$CO$^+$ model can be used
  for coadding faint emission lines and obtain a sufficient signal-to-noise ratio to detect
  this isotopologue.}
   {Our fluorescence model provides a good modeling of the $^{12}$CO$^+$
   emission lines, allowing to publish revised fluorescence efficiencies. Based on similar
   transition probabilities for $^{12}$CO$^+$ and $^{13}$CO$^+$ we derive a 
   $^{12}$C/$^{13}$C isotopic ratio of 73$\pm$20 for CO$^+$ in comet C/2016 R2. This value is 
   in agreement with the solar system ratio of 89$\pm$2 within the error bars, making the possibility 
   that this comet was an interstellar object unlikely.
   }
   {}


   \maketitle
%
%-------------------------------------------------------------------

\section{Introduction}

Comets are small icy bodies that remain relatively unaltered by physico-chemical
processes since their formation in the outer part of the Solar System. Studying their
physical and chemical properties provides useful constraints on the physical and 
chemical properties of their formation place and, consequently, on the protosolar 
nebula. These small bodies present some significant differences in their chemical
composition but their main species are known for a long time thanks to numerous spectroscopic
observations. The main species detected in cometary coma is water molecules with 
CO and CO$_2$ being the second most abundant species (typically about 10–20\% relative to water).

Carbon monoxide being much more volatile than water this species can drive 
cometary activity at large distance from the Sun (heliocentric distance larger than 5~au)
while water is responsible for the cometary activity usually observed, i.e. for comets closer
than about 3~au from the Sun. In the near-UV and optical range, where most of spectroscopic
observations are conducted, it is possible to observe CO$^+$ emission lines, this ion being
created by CO, or indirectly by CO$_2$. This species has been first reproduced in the 
laboratory by \cite{fowler:1909a,fowler:1909b} for spectra obtained in the tails of comets 
C/1907 L2 (Daniel) and C/1908 R1 (Morehouse) and later assigned to CO$^+$. The same author later 
noticed the presence of similar emission bands in Brorsen's comet observed in 1868 by 
William Huggins \citep{fowler:1910}. \cite{swings:1965} mentions observations of CO$^+$ bands 
(now called "comet tail system of CO$^+$") in the tail of several comets observed after
C/1908 R1 (Morehouse).

Emission bands of CO$^+$ are, nevertheless, not so often observed in comets. As mentioned
above such bands were first observed in the tails of comets at an epoch where 
spectrographs observed comets at a large scale, covering both the coma and the tail. 
The instruments later focused on the inner coma, where the signal-to-noise is better and
permits to get high-resolution spectra. In this region the detection of CO$^+$ species is more 
difficult. A noticeable exception was the comet C/2016 R2 (PanSTARRS) whose coma presented
an unusual composition dominated by the N$_2^+$ and CO$^+$ ions \citep{cochran:2018a,opitom:2019}.

This comet was discovered on September 7, 2016 by the PanSTARRS survey \citep{weryk:2016}.
It is a long period comet that developed a coma as far as 6~au from the Sun
and started to display unusual coma morphology with structure changing rapidly\footnote{See
https://www.eso.org/public/videos/potw1940a/}, attributed
to ion dominating the emission in the coma. Radio observations revealed a coma dominated
by CO with a low abundance of HCN \citep{Wierzchos:2018} and optical observations
later revealed a spectrum dominated by CO$^+$ and N$_2^+$ emission bands 
\citep{Biver:2018,cochran:2018a,opitom:2019}. The unusually high abundance of CO$^+$ ion in the inner
coma of this relatively bright comet (and unusually low abundance of common species like CN and C$_2$)
permitted to get CO$^+$ spectra with unprecedented quality
(both with high-resolution and high signal-to-noise ratio) for this species with the UVES
spectrograph mounted on the 8.2-m ESO VLT telescope, already presented in \cite{opitom:2019}.

A correct analysis of such spectra implies a good modeling of the fluorescence spectrum 
of CO$^+$ in comets. Such a modeling has already been done by \cite{magnani:1986}. Nevertheless
this work does not present any spectrum that could be confronted to our observational data
and some new laboratory and
theoretical works have been published after this pioneering work. For these reasons we decided
to build a new fluorescence model, based on more recent laboratory data and theoretical
works.

In this paper we first present our observational data before explaining in details 
our model and compare it to the observed spectra. This work also provides new information related 
to the fluorescence efficiencies. Because such data also open the possibility, for the first time, 
to measure the $^{12}$C/$^{13}$C isotopic ratio in CO$^+$ we also computed a fluorescence 
model of the $^{13}$CO$^+$ isotopologue and searched for its emission lines. The result of the
search of these faint lines is also presented, leading to a first estimate of the
$^{12}$CO$^+$/$^{13}$CO$^+$ ratio in a comet.

\section{Observational data}

The spectra used for this work have been obtained with the Ultraviolet-Visual Echelle Spectrograph
(UVES) mounted on the ESO 8.2 m UT2 telescope of the Very Large Telescope (VLT) located in Chile. 
They correspond to the dichroic~\#1 (390+580 settings) covering the range 326 to 454~nm 
in the blue and 476 to 684~nm in the red
and the dichroic~\#2 (437+860 settings) covering the range 373 to 499~nm in the blue and 660 to
1060~nm in the red (the data obtained in this last spectral range have not been used for this work).
Three different observing nights have been used for observations with the dichroic \#1, 
corresponding to February 11, 13 and 14, 2018. During each night one single exposure of 4800~s of 
integration time was obtained. We used a 0.44'' wide slit, providing a resolving power of 
R$\sim$80,000. The slit length was 8'' for the setting 390 corresponding to about 14,500 km at the 
distance of the comet (geocentric distance of 2.4 au) and 12'' for the setting 580. 
The average heliocentric distance was 2.76~au and the heliocentric velocity 5.99~km~s$^{-1}$.
The observations performed with dichroic \#2 (437+860) correspond to two observing nights
on February 15 and 16. In that case a single exposure of 3000~s of integration time was
obtained for both nights and the slit width was also 0.44'' with a slit length of 10'' (for the 
setting 437). \cite{opitom:2019} provides more details about these observations.

As explained by these authors the data were reduced using the ESO UVES pipeline, combined 
with custom routines to perform the extraction, cosmic rays removal, and then corrected for the 
Doppler shift due to the relative velocity of the comet with respect to the Earth. The spectra 
are calibrated in absolute flux using either the archived master response curve or the response curve 
determined from a standard star observed close to the science spectrum (with no significant differences between these two methods). This data processing produced 2D spectra calibrated in
wavelength and absolute flux units. Because a close examination of the ESO UVES sky emission 
spectrum\footnote{https://www.eso.org/observing/dfo/quality/UVES/pipeline/ sky\_spectrum.html}
permits to check that telluric lines are below the noise level in this part of the spectrum no 
specific data processing was done for removing these lines.

From these three different settings (390, 580 and 437) we computed average 1D spectra. These
1D spectra were further splitted in different parts corresponding to the different
CO$^+$ emission bands. The solar continuum has been subtracted for each of these different
spectra individually in order to adjust it as well as possible in a limited spectral range.
Many CO$^+$ bands can be identified in our observational data (see section 5 for more details).

%Table~3 of \cite{opitom:2019} provides an overview of the different CO$^+$ bands detected 
%as well as their wavelengths.

\section{$^{12}$CO$^+$ fluorescence model}

\begin{figure*}
\centering
\includegraphics[width=12cm]{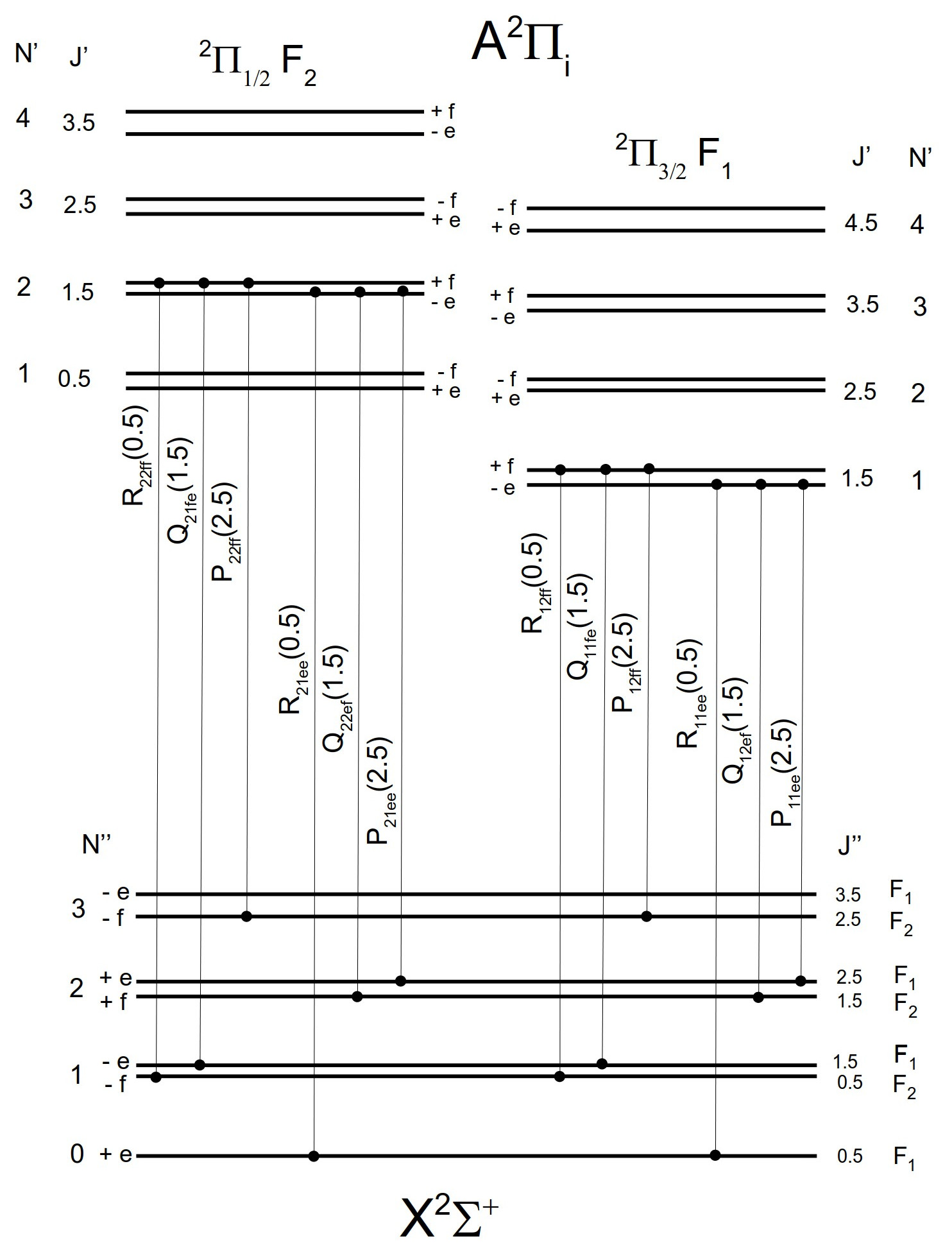}
\caption{Energy level diagram showing the A$^2\Pi_i$--X$^2\Sigma^+$ transition
with the different lines involved by this transition. The number in parenthesis refers
to the $J''$ value of the line.}
\label{figAX}
\end{figure*}

The CO$^+$ emission lines observed in the optical range belong to the comet tail system, i.e. to the
A$^2\Pi_i$--X$^2\Sigma^+$ electronic transition. The A$^2\Pi_i$ state is divided into two 
branches ($^2\Pi_{1/2}$ and $^2\Pi_{3/2}$ labeled $F_2$ and $F_1$) with a large energy separation
because the $2\Pi$ electronic state is intermediate between Hund's case (a) and (b) (the $F_2$ states 
corresponding to higher energies). The rotational levels are split as a result of 
$\Lambda$-doubling. Such a rotational structure gives rise to 12 branches and each band
is divided into two sub-bands corresponding to the $^2\Pi_{1/2}$--$X^2\Sigma^+$ and 
$^2\Pi_{3/2}$--$X^2\Sigma^+$ transitions. Fig. \ref{figAX} presents the 
energy levels diagram with the different types of lines involved in this transition.

A pioneering work was published by \cite{magnani:1986} that presents a fluorescence model
for CO$^+$ in cometary coma. This
paper does not show any comparison between synthetic spectra and observational data. 
For modeling the $^{12}$CO$^+$ and $^{13}$CO$^+$ in comet C/2016 R2 it appears necessary
to develop a new fluorescence model that can benefit from different
experimental and theoretical articles published after Magnani \& A'Hearn's paper. Such 
works can significantly help to improve this modeling. \cite{magnani:1986} contains nevertheless
interesting data to built such a new fluorescence model.

We developed our own fluorescence model of CO$^+$ that takes into account the A$^2\Pi_i$ and
X$^2\Sigma^+$
electronic levels with the first 6 vibronic levels ($v=0$ to $5$).
The energy levels for the X$^2\Sigma^+$ state and the first three vibrational quantum
numbers $v$ equal to 0, 1 and 2 have been taken from \cite{hakalla:2019}, as well as the
$v=0,1$ vibronic energy levels of the A$^2\Pi_i$ state. From the X$^2\Sigma^+$ $v=0$
levels accurate energy levels of the A$^2\Pi_i$ $v=2,3,4$ states have been computed
with the transition frequencies published by \cite{kepa:2004} for the (2,0), (3,0) and
(4,0) bands. The remaining energy levels ($v=3,4,5$ of the X$^2\Sigma^+$ state and
$v=5$ of the A$^2\Pi_i$ state) have been computed from the molecular parameters 
published by \cite{kepa:2004} (rotational constants) and \cite{coxon:2010} (band origin).
All the levels with a rotational quantum number $J\leq10.5$ have been
taken into account in our model.

The Einstein coefficients for the spontaneous transition probabilities have been 
computed with the transition probabilities computed by \cite{billoux:2014} 
expressed in atomic units. These values need to be converted to spontaneous emission
Einstein coefficients $A_{ul}$ from the upper level $u$ to the lower level $l$ 
of a given line by using the following formula (case of a $\Pi$--$\Sigma$ transition):

$$A_{ul}=2.026\times 10^{-6}\times \sigma_{ul}^3\times p^{v'v''}_{N'N''}{S_{J'J''}\over 2J'+1}$$

with $p^{v'v''}_{N'N''}$ being the transition probability provided by \cite{billoux:2014}
expressed in atomic units, $\sigma_{ul}$ the wavenumber expressed in cm$^{-1}$,
$J'$ the rotational quantum number of the upper state and $S_{J'J''}$ the H\"onl-London
factor.

The H\"onl-London factors have been taken from \cite{arpigny:1964a} and renormalized
to follow the summation rule used by \cite{billoux:2014}, i.e.:

$$\sum_{J'}S_{J'J''}=(2-\delta_{0,\Lambda'+\Lambda''})(2S'+1)(2J''+1)$$

The Einstein absorption coefficients $B_{lu}$ have been computed from the $A_{ul}$ coefficients and 
the wavenumbers of the corresponding transitions.
The probability of absorption is given by $B_{lu}\times \rho_\nu$ 
with $\rho_\nu$ being the radiation density at the corresponding wavelength, expressed in 
erg.cm$^{-3}$.Hz$^{-1}$. We used the high-resolution 
solar spectrum published by \cite{kurucz:1984} to compute the solar radiation density.

The pure vibrational transition probabilities of the ground X$^2\Sigma^+$ state have 
been taken in \cite{rosmus:1982} (A$_{v'v''}$ Einstein coefficients) with H\"onl-London
factors from \cite{magnani:1986}. The pure rotational transition probabilities have been computed 
from the formulae published by \cite{arpigny:1964b} for a $^2\Sigma$ state in using an electric dipole
moment $\mu=2.771$~Debye \citep{bell:2007}.

Apart the ground electronic state $X^2\Sigma^+$ and the first electronic excited state
$A^2\Pi_i$ there is another excited electronic state $B^2\Sigma^+$ with a $B^2\Sigma^+$--$X^2\Sigma^+$
transition called first negative system and a $B^2\Sigma^+$--$A^2\Pi_i$ transition called
Baldet-Johnson system (see Fig. 1 in \cite{magnani:1986}). Because the first negative system 
B$^2\Sigma^+$--X$^2\Sigma^+$ transition can 
also slightly influence the relative populations of the ground electronic state some levels in the 
B$^2\Sigma^+$ ($v=0,1,2$) have also being taken into account in our modeling with the first 
negative system (the Baldet-Johnson system implying two excited electronic states its influence
can be neglected).
The energy levels of the $B^2\Sigma^+$ have been computed with the molecular parameters 
published by \cite{Szajna:2004} and the transition probabilities are taken from \cite{magnani:1986}.

The relative population have been computed at the equilibrium with the method described by 
\cite{zucconi:1985}. Such a method is justified by the heteronuclear nature of the CO$^+$ ion
that involves pure vibrational and rotational transitions that permit to reach the fluorescence
equilibrium with a timescale well below the time spent by these ions in the observed cometary
coma (this is not the case for homonuclear species like, e.g. C$_2$ or N$_2^+$). We assume that the 
coma is an optically thin medium for this spectrum, i.e. that the 
radiation received by the telescope is proportional to the number of CO$^+$ ions along the line of 
sight with the radiation emitted in units of energy in $4\pi$ steradian equal to $x_i\times h\nu$, $x_i$
being the relative population, $h$ the Planck's constant and $\nu$ the frequency.

\section{$^{13}$CO$^+$ fluorescence model}

For this isotopologue the energy levels comes from the laboratory data published by 
\cite{kepa:2002}: we used the (1,0), (3,0), (4,0) and (5,0) bands wavenumbers 
of the lines observed to 
reconstruct all the energy levels of the X$^2\Sigma^+$ state $v=0$. From this state it was then 
possible to compute the energy levels of the A$^2\Pi_i$ $v=1,2,3,4,5$ states by using
the (1,0)(2,0)(3,0)(4,0)(5,0) transitions. After this computation the X$^2\Sigma^+$ state $v=1$ 
was computed by using the (2,1) band wavenumbers.
Such a method provide a high accuracy for the final wavelength (accuracy of about 0.01~\AA, well 
below the FWHM of our high-resolution spectra). The missing energy levels were computed from the 
molecular parameters published by \cite{kepa:2002} for the A$^2\Pi_i$ $v=0$ state and X$^2\Sigma^+$ 
state $v=2,3,4,5$.

For the B$^2\Sigma^+$ level we also used the molecular constants published by \cite{Szajna:2004} but
adapted for the $^{13}$CO$^+$ isotopologue. It was done by using the reduced atomic mass $\mu'$ of 
this isotopologue and the one of $^{12}$CO$^+$ ($\mu$) providing the parameter 
$\rho=\sqrt{\mu/\mu'}=0.9777$. This parameter was used to compute the molecular parameters of
$^{13}$CO$^+$ from the ones of $^{12}$CO$^+$ thanks to the formulae given 
by \cite{Herzberg:1950}.

The transition probabilities have been assumed to be equal to the ones of $^{12}$CO$^+$, such an 
approximation is justified by some calculations done for other diatomic species that show
only small relative differences for the transition probabilities for different isotopologues
(see, e.g., \cite{ferchichi:2022} for the different isotopologues of N$_2^+$).

\section{Comparison with observational data and fluorescence efficiencies}

The $^{12}$CO$^+$ emission lines being clearly visible in our C/2016 R2 spectra it is possible to 
check, for the first time, with a high accuracy the quality of our fluorescence model 
developed for this isotopologue. Our observational data permit to observe different bands
with the three different UVES settings mentioned in section 2. These settings corresponding to different
slit sizes, epochs of observations or exposure times they show different relative intensities of the spectra.

\begin{figure*}
\centering
\includegraphics[width=18cm]{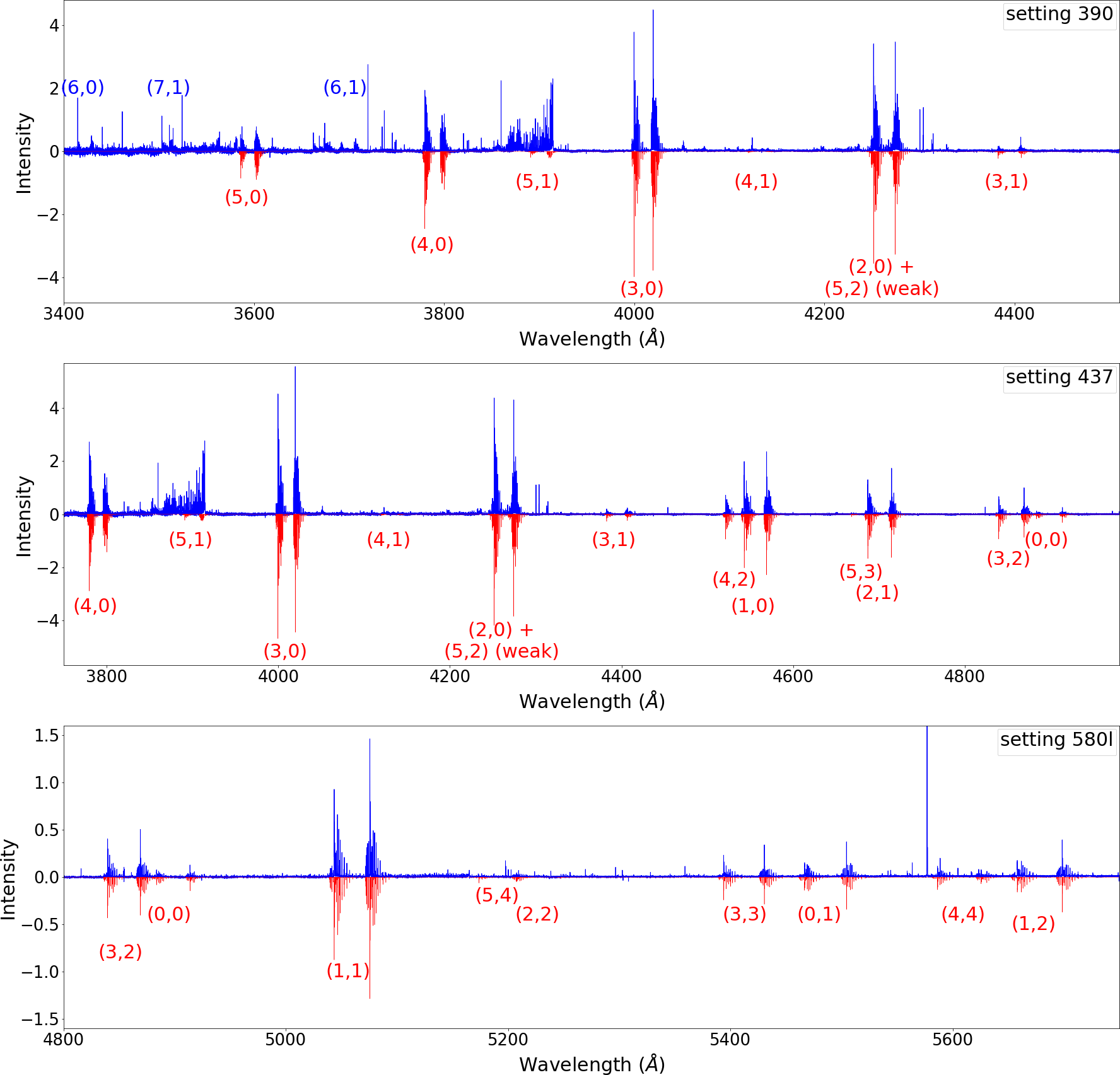}
\caption{Comparison of our modeling of $^{12}$CO$^+$ spectrum with our observational data obtained
on comet C/2016 R2 for the three different settings. The observational spectra appear in blue while
the model is in red (for more clarity it appears with a negative scale). The different CO$^+$
bands bright enough to be detected in the observational spectrum are identified in red for the 
model. 
%Some other species that appear in the observational data are identified in blue. 
The (6,0), (7,1) and (6,1) CO$^+$ bands appear (weakly) 
on the observational spectrum and are identified in blue, but not on the model because this one is
restricted to $v''$ and $v'\leq 5$. The intensity scale is arbitrary but proportional to 
units of ergs.s$^{-1}$. Some other species appear in the observational spectrum (e.g. N$_2^+$
and CN emission lines near the (5,1) CO$^+$ band).}
\label{overallfit}
\end{figure*}

We computed the $^{12}$CO$^+$ fluorescence spectrum for the condition of observations of
C/2016 R2 and adjusted its intensity for each setting. We kept the same relative 
intensity for all the different bands observed with a same setting, in order to check the
accuracy of the relative intensities provided by our model for different bands.
Fig.~\ref{overallfit} provides an overview of all the CO$^+$ emission bands identified 
in our observations, for the three different settings used for this work.

\begin{figure*}
\centering
\includegraphics[width=18cm]{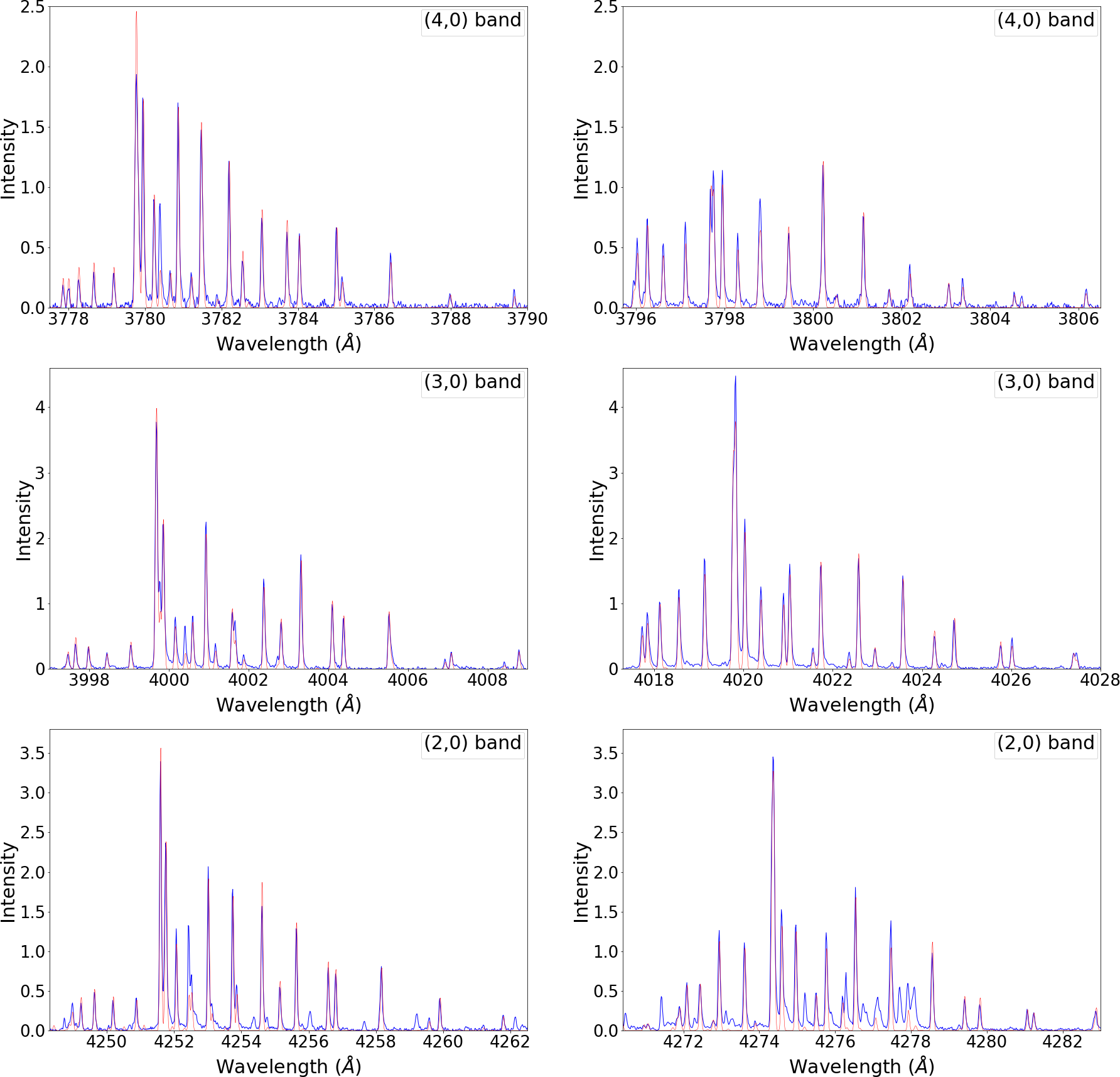}
\caption{Comparison of our modeling of $^{12}$CO$^+$ spectrum with our observational data obtained
on comet C/2016 R2, for different bands observed with the setting 390. The intensity scale is 
arbitrary but proportional to units of ergs.s$^{-1}$ and is the same for the different bands. 
The observational spectrum appears in blue while the modeling is in red. The emission lines
of the (2,0) band are blended with some emission lines of the (5,2) band but this band
represents only a few percent in intensity with respect to the (2,0) band.}
\label{fit390}
\end{figure*}

\begin{figure*}
\centering
\includegraphics[width=18cm]{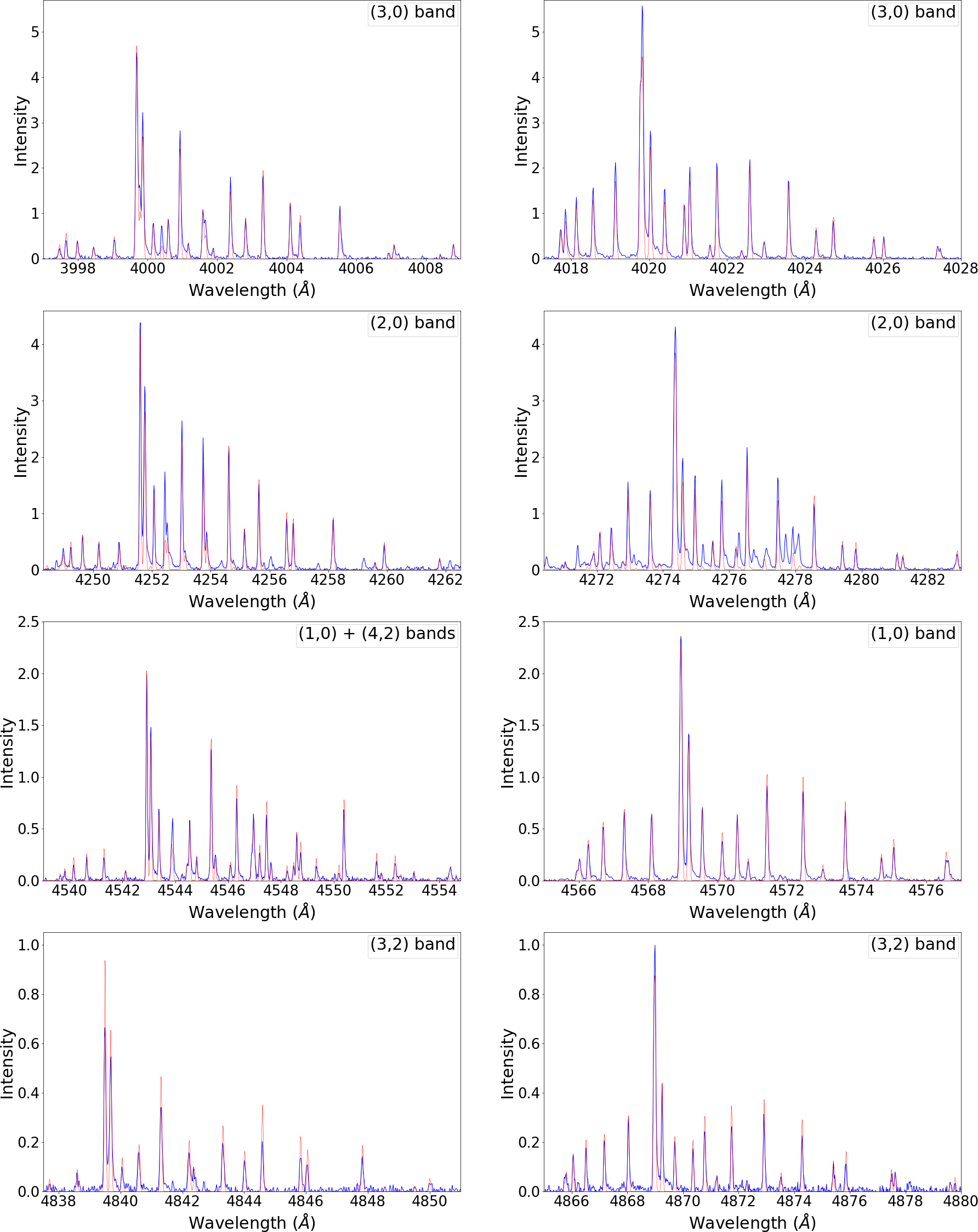}
\caption{Comparison of our modeling of $^{12}$CO$^+$ spectrum with our observational data obtained
on comet C/2016 R2, for different bands observed with the setting 437. The intensity scale is arbitrary 
but proportional to units of ergs.s$^{-1}$ and is the same for the different bands. The observational 
spectrum appears in blue while the modeling is in red. The emission lines of (1,0) $\Pi_{1/2}$ band are
mixed with the emission lines of the (4,2) $\Pi_{3/2}$ band, these last ones being fainter (intensities
roughly equal to the ones of the (3,2) $\Pi_{3/2}$ band).}
\label{fit437}
\end{figure*}

\begin{figure*}
\centering
\includegraphics[width=18cm]{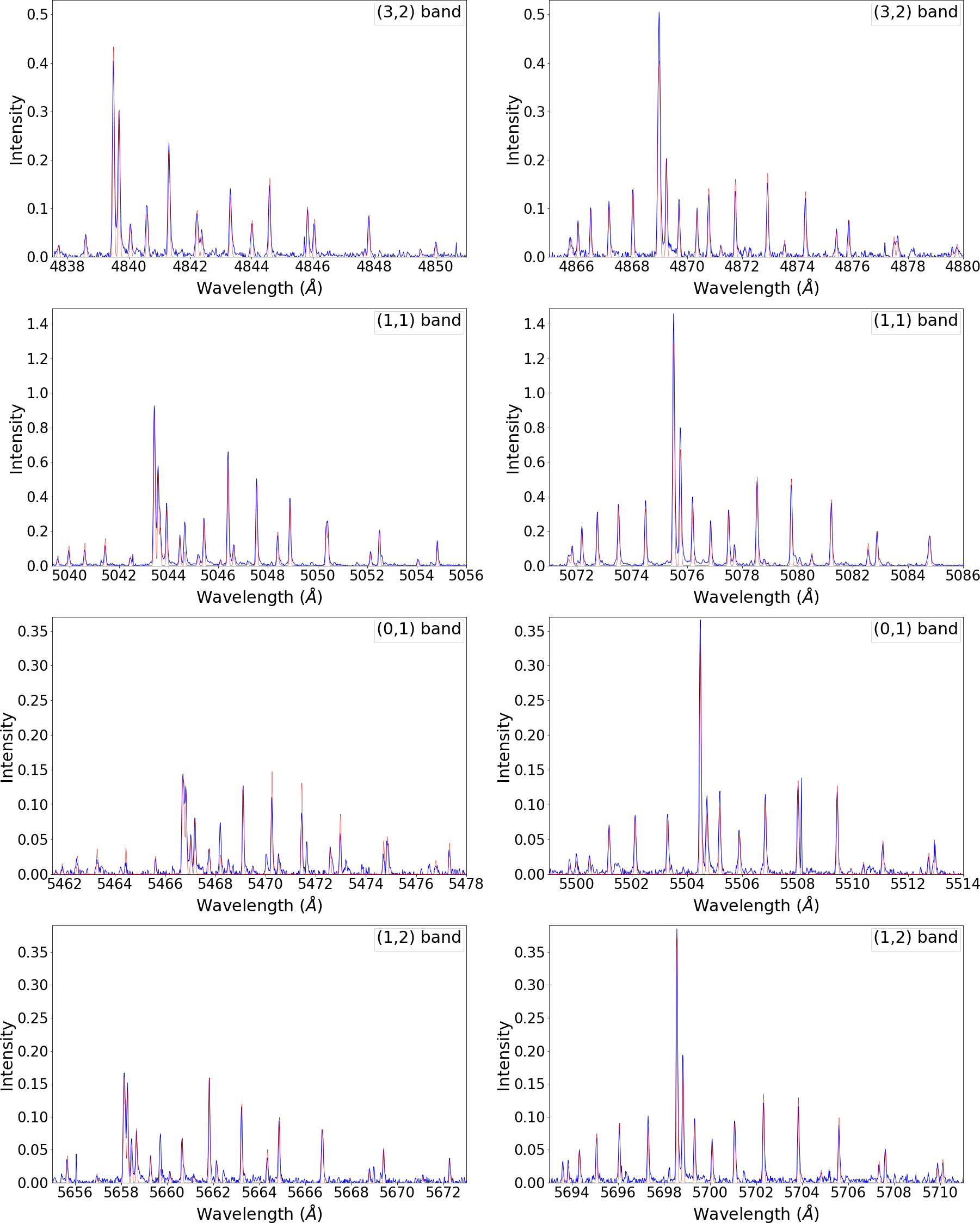}
\caption{Comparison of our modeling of $^{12}$CO$^+$ spectrum with our observational data obtained
on comet C/2016 R2, for different bands observed with the setting 580l. The intensity scale 
is arbitrary but proportional to units of ergs.s$^{-1}$ and is the same for the different bands. 
The observational spectrum appears in blue while the modeling is in red.}
\label{fit580l}
\end{figure*}

For a better comparison between observational data and the synthetic spectrum Fig.~\ref{fit390}, 
Fig.~\ref{fit437} and Fig.~\ref{fit580l} zooms on different bright bands to show the details
of the rotational structure. The 
relative intensities for the different bands agree well with the observational data.

\begin{figure*}
\centering
\includegraphics[width=15cm]{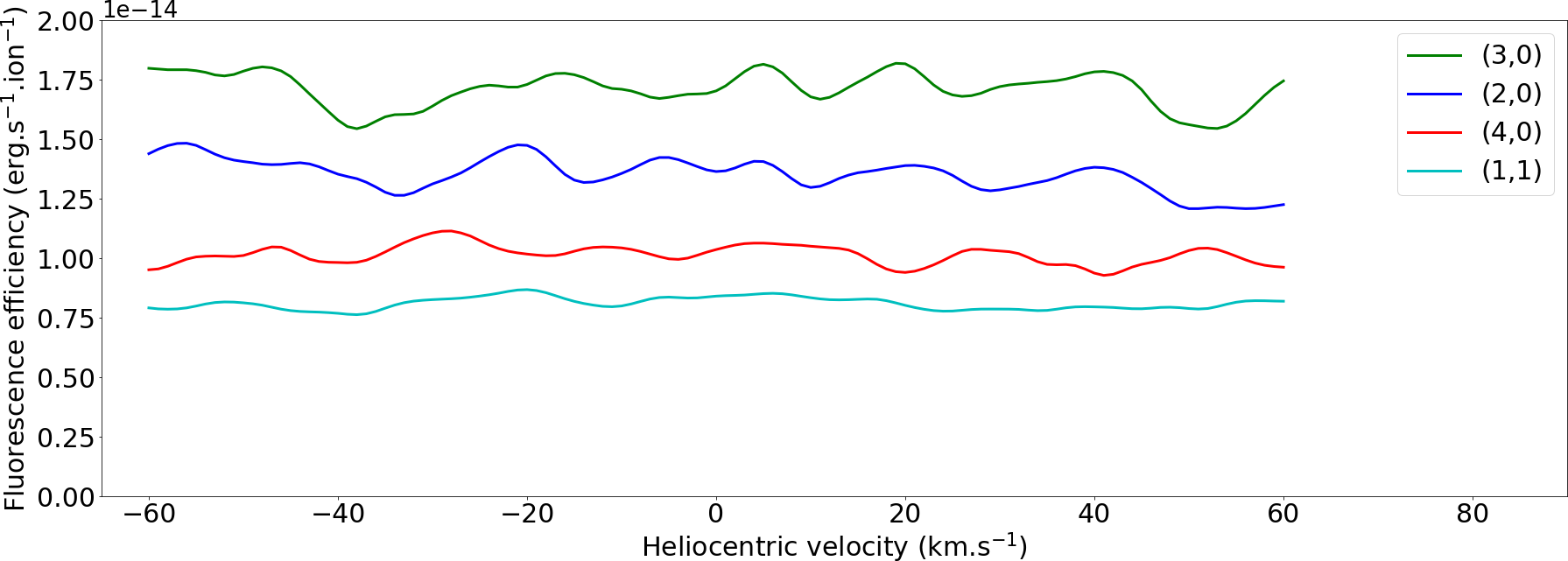}
\caption{Fluorescence efficiencies of the four brightest CO$^+$ bands as a function of the heliocentric velocity.}
\label{g_rh}
\end{figure*}

We can compute fluorescence efficiencies, also called $g$-factors or band luminosities 
with our fluorescence model. Table \ref{tab:gfactors1ua} provides an overview of these important 
parameters for an heliocentric distance of 1~au and an heliocentric velocity of  
0~km.s$^{-1}$. A comparison of this parameter with the results published by \cite{magnani:1986} in 
their Table 3 permits to see a similar value for the (4,0) band ($1.07\times 10^{-14}$ vs $1.04\times 
10^{-14}$~erg.s$^{-1}$.ion$^-1$), which is one of the brightest band. The relative differences in intensities 
between our model and the results published by \cite{magnani:1986}
can reach a factor of about two for some bands. For the brightest ones (i.e. (3,0), (2,0) and (4,0) 
by decreasing intensity), that are displayed Fig. \ref{fit390} (and Fig. \ref{fit437} for (2,0) and (3,0)) 
our modeling shows a good fit with the observational data. Our computed intensity ratio
$g(3,0)/g(4,0)=1.63$ appears in 
good agreement with the observational data but disagrees with the ratio computed by 
\cite{magnani:1986}, which is $2.03$ and that would not be able to fit such observational data 
as well as our model. We have no possibility to test the absolute values of the band luminosities but 
our fit of the different bands confirms that their relative values are robust, at least to about 10\%.
Fig. \ref{g_rh} shows the variation of fluorescence efficiencies as a function of the heliocentric 
velocity for the four brightest bands (i.e. (3,0), (2,0), (4,0) and (1,1) by decreasing intensity). 
Some variations can be observed but they remain limited to less than about 15\% between the smaller 
and the larger values. We also computed the product of the fluorescence efficiency by the square 
of heliocentric distance $R_h$ (expressed in au) in order to check if there is any significant 
deviation from the usual scaling law with this parameter. Only some small variations can be observed (see 
Table~\ref{tab:gfactorsRh}).

\begin{table*}
  \caption{Fluorescence efficiencies computed for $^{12}$CO$^+$, for the heliocentric distance 
  $r_h=1$~au and the heliocentric velocity $v_h=0$~km.s$^{-1}$, in units of 
  erg.s$^{-1}$.ion$^{-1}$.}
  \centering 
	\centering
    \begin{tabular}{ccccccc}
    \hline
    %{m{15mm} m{70mm} m{18mm}}
   & $v"=0$ & 1 & 2 & 3 & 4 & 5\\
    $v'$\\
     \hline
0 &  8.40$\times 10^{-16}$  & 1.87$\times 10^{-15}$  & 1.81$\times 10^{-15}$  & 9.99$\times 10^{-16}$  & 3.45$\times 10^{-16}$  & 7.63$\times 10^{-17}$  \\
1 &  7.68$\times 10^{-15}$  & 8.41$\times 10^{-15}$  & 2.19$\times 10^{-15}$  & 7.10$\times 10^{-18}$  & 7.63$\times 10^{-16}$  & 7.87$\times 10^{-16}$  \\
2 &  1.36$\times 10^{-14}$  & 5.38$\times 10^{-15}$  & 9.28$\times 10^{-17}$  & 2.25$\times 10^{-15}$  & 1.03$\times 10^{-15}$  & 2.94$\times 10^{-18}$  \\
3 &  1.70$\times 10^{-14}$  & 9.24$\times 10^{-16}$  & 3.07$\times 10^{-15}$  & 1.90$\times 10^{-15}$  & 5.49$\times 10^{-17}$  & 8.79$\times 10^{-16}$  \\
4 &  1.04$\times 10^{-14}$  & 1.97$\times 10^{-16}$  & 2.96$\times 10^{-15}$  & 2.23$\times 10^{-17}$  & 9.53$\times 10^{-16}$  & 2.80$\times 10^{-16}$  \\
5 &  5.12$\times 10^{-15}$  & 1.21$\times 10^{-15}$  & 1.04$\times 10^{-15}$  & 4.14$\times 10^{-16}$  & 5.21$\times 10^{-16}$  & 4.16$\times 10^{-17}$  \\
    \hline
    \end{tabular}
\label{tab:gfactors1ua}
  \end{table*}

\begin{table*}
  \caption{Fluorescence efficiencies multiplied by $R_h^2$ computed for $^{12}$CO$^+$.}
  \centering 
	\centering
    \begin{tabular}{ccccccc}
    \hline
    %{m{15mm} m{70mm} m{18mm}}
   & $R_h=0.5$~au & $1.0$ & $1.5$ & $2.0$ & $3.0$ & $4.0$\\
    band\\
     \hline
(2,0) &  1.35$\times 10^{-14}$  & 1.36$\times 10^{-14}$   &  1.37$\times 10^{-14}$   &  1.38$\times 10^{-14}$  &  1.39$\times 10^{-14}$   &  1.41$\times 10^{-14}$  \\

(3,0) &  1.67$\times 10^{-14}$  &  1.70$\times 10^{-14}$  &  1.72$\times 10^{-14}$   &  1.74$\times 10^{-14}$  &  1.77$\times 10^{-14}$  &  1.79$\times 10^{-14}$  \\

(4,0) &  1.02$\times 10^{-14}$   & 1.04$\times 10^{-14}$    &  1.04$\times 10^{-14}$    &  1.03$\times 10^{-14}$   &  1.01$\times 10^{-14}$   &  9.94$\times 10^{-15}$ \\

(1,1) &  8.28$\times 10^{-15}$   &  8.41$\times 10^{-15}$  &  8.46$\times 10^{-15}$  &  8.47$\times 10^{-15}$  &  8.47$\times 10^{-15}$  &  8.43$\times 10^{-15}$   \\

    \hline
    \end{tabular}
\label{tab:gfactorsRh}
  \end{table*}

We also computed fluorescence efficiencies for $^{13}$CO$^+$ (see Table~\ref{tab:gfactors1ua13co}). 
As we assumed that $^{13}$CO$^+$ and $^{12}$CO$^+$ have similar transition probabilities these 
band luminosities are very close to the ones of $^{12}$CO$^+$ but some differences appear due to 
the wavelength shifts that implies some variations in the solar flux for absorption transitions. 
Such fluorescence efficiencies can be useful for future $^{12}$C/$^{13}$C isotopic ratio measurements in CO$^+$ (see below).

\begin{table*}
  \caption{Fluorescence efficiencies computed for $^{13}$CO$^+$, for the heliocentric distance 
  $r_h=1$~au and the heliocentric velocity $v_h=0$~km.s$^{-1}$, in units of erg.s$^{-1}$.ion$^{-1}$.}
  \centering 
	\centering
    \begin{tabular}{ccccccc}
    \hline
    %{m{15mm} m{70mm} m{18mm}}
   & $v"=0$ & 1 & 2 & 3 & 4 & 5\\
    $v'$\\
     \hline

0  & 8.89$\times 10^{-16}$  & 1.99$\times 10^{-15}$  & 1.96$\times 10^{-15}$  & 1.10$\times 10^{-15}$  & 3.88$\times 10^{-16}$  & 8.85$\times 10^{-17}$  \\
1  & 6.84$\times 10^{-15}$  & 7.56$\times 10^{-15}$  & 1.99$\times 10^{-15}$  & 6.55$\times 10^{-18}$  & 7.16$\times 10^{-16}$  & 7.55$\times 10^{-16}$  \\
2  & 1.38$\times 10^{-14}$  & 5.49$\times 10^{-15}$  & 9.56$\times 10^{-17}$  & 2.34$\times 10^{-15}$  & 1.09$\times 10^{-15}$  & 3.17$\times 10^{-18}$  \\
3  & 1.65$\times 10^{-14}$  & 9.03$\times 10^{-16}$  & 3.02$\times 10^{-15}$  & 1.89$\times 10^{-15}$  & 5.53$\times 10^{-17}$  & 8.98$\times 10^{-16}$  \\
4  & 7.69$\times 10^{-15}$  & 1.47$\times 10^{-16}$  & 2.23$\times 10^{-15}$  & 1.68$\times 10^{-17}$  & 7.29$\times 10^{-16}$  & 2.17$\times 10^{-16}$  \\
5  & 5.40$\times 10^{-15}$  & 1.28$\times 10^{-15}$  & 1.11$\times 10^{-15}$  & 4.46$\times 10^{-16}$  & 5.66$\times 10^{-16}$  & 4.57$\times 10^{-17}$  \\

    \hline
    \end{tabular}
\label{tab:gfactors1ua13co}
  \end{table*}

\section{Measurement of $^{12}$CO$^+$/$^{13}$CO$^+$ isotopic ratio in comet C/2016 R2}

\begin{figure*}
\centering
\includegraphics[width=17cm]{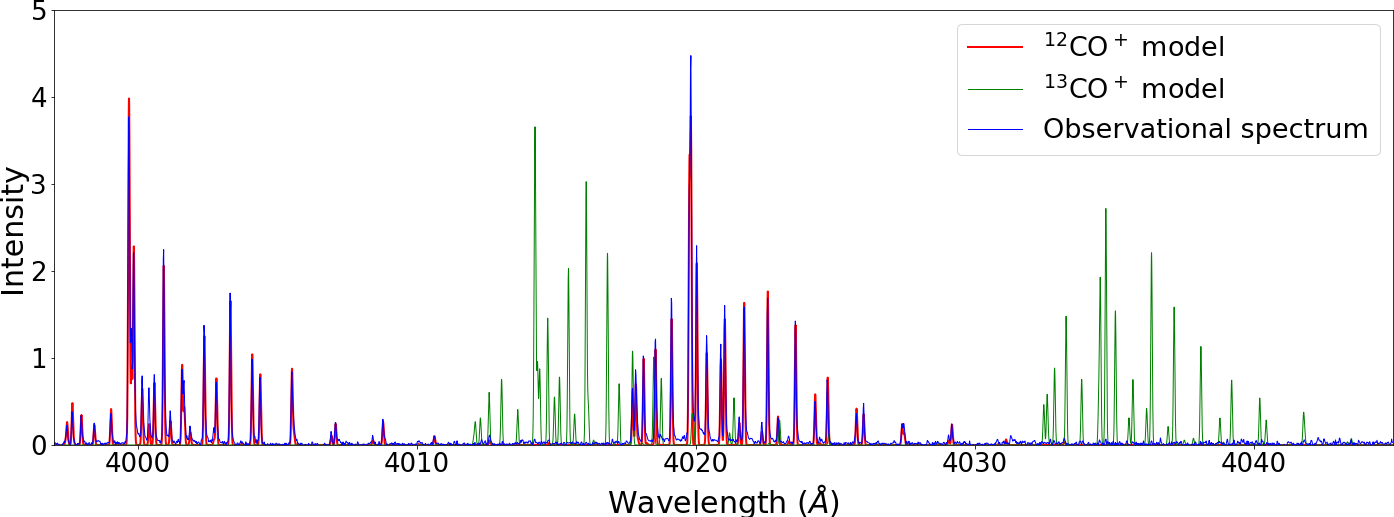}
\caption{$^{12}$CO$^+$ and $^{13}$CO$^+$ theoretical spectra superimposed to the observational data
obtained on comet C/2016 R2. The relative intensity of $^{13}$CO$^+$ emission lines are based
on the $^{12}$CO$^+$ fit of the emission lines. They do not respect the intensities
observed in this comet, the purpose of this plot being only to show the wavelength shift between
the emission lines of these two isotopologues.}
\label{band_30_12CO_13CO}
\end{figure*}

The high signal-to-noise and spectral resolution of our observational data opens the possibility, 
for the first time, to measure the $^{12}$C/$^{13}$C isotopic ratio in the CO$^+$ species with ground-based
observations. It is 
possible thanks to the wavelength shift between the spectrum of the two isotopologues. 
Fig.~\ref{band_30_12CO_13CO} presents the synthetic spectra of these two isotopologues
superimposed to the observational data of the brightest band (i.e. the (3,0)). It can be seen that 
the $^{13}$CO$^+$ emission lines are clearly separated of the $^{12}$CO$^+$ ones. At this intensity 
scale no $^{13}$CO$^+$ emission lines are visible.

To increase the signal-to-noise ratio of the possible $^{13}$CO$^+$ emission lines we decided to 
coadd the brightest $^{13}$CO$^+$ emission lines for the two brightest bands, i.e. the (3,0) and 
(2,0). The wavelengths have been carefully chosen to avoid any other possible emission lines due to 
other species. At the end we restricted our choice to 15 lines belonging to the (3,0) band and 9 
lines for the (2,0) band. Their counterpart for the $^{12}$CO$^+$ were also coadded. 
Table~\ref{tab:lines12C13Clist} provides the details of the lines wavelengths used 
as well as their $^{12}$CO$^+$ counterparts. The work here was facilitated because C/2016 R2 
is very poor in the usual bright and abundant lines of CN and C$_2$ in the CO$^+$ regions of interest.

\begin{table}
  \caption{Wavelengths ({\AA}) of the $^{13}$CO$^+$ lines coadded to measure the
  $^{12}$C/$^{13}$C isotopic ratio with their $^{12}$CO$^+$ counterpart.}
  \centering 
	\centering
    \begin{tabular}{ccc}
    \hline
    %{m{15mm} m{70mm} m{18mm}}
   Band & $^{13}$CO$^+$ & $^{12}$CO$^+$ counterpart\\
     \hline

(3,0) & 4032.600 & 4017.855 \\
(3,0) & 4032.865 & 4018.135 \\
(3,0) & 4033.275 & 4018.560 \\
(3,0) & 4033.835 & 4019.140 \\
(3,0) & 4034.500 & 4019.825 \\
(3,0) & 4034.700 & 4020.040 \\
(3,0) & 4035.045 & 4020.395 \\
(3,0) & 4035.670 & 4021.045 \\
(3,0) & 4036.340 & 4021.740 \\
(3,0) & 4036.930 & 4022.375 \\
(3,0) & 4038.790 & 4024.285 \\
(3,0) & 4039.210 & 4024.730 \\
(3,0) & 4040.215 & 4025.765 \\
(3,0) & 4040.445 & 4026.020 \\
(3,0) & 4041.790 & 4027.402 \\
(2,0) & 4283.360 & 4272.445 \\
(2,0) & 4283.840 & 4272.950 \\
(2,0) & 4284.480 & 4273.620 \\
(2,0) & 4285.200 & 4274.370 \\
(2,0) & 4285.425 & 4274.600 \\
(2,0) & 4285.780 & 4274.970 \\
(2,0) & 4286.295 & 4275.510 \\
(2,0) & 4288.195 & 4277.475 \\
(2,0) & 4290.445 & 4279.825 \\

    \hline
    \end{tabular}
\label{tab:lines12C13Clist}
  \end{table}

\begin{figure*}
\centering
\includegraphics[width=12cm]{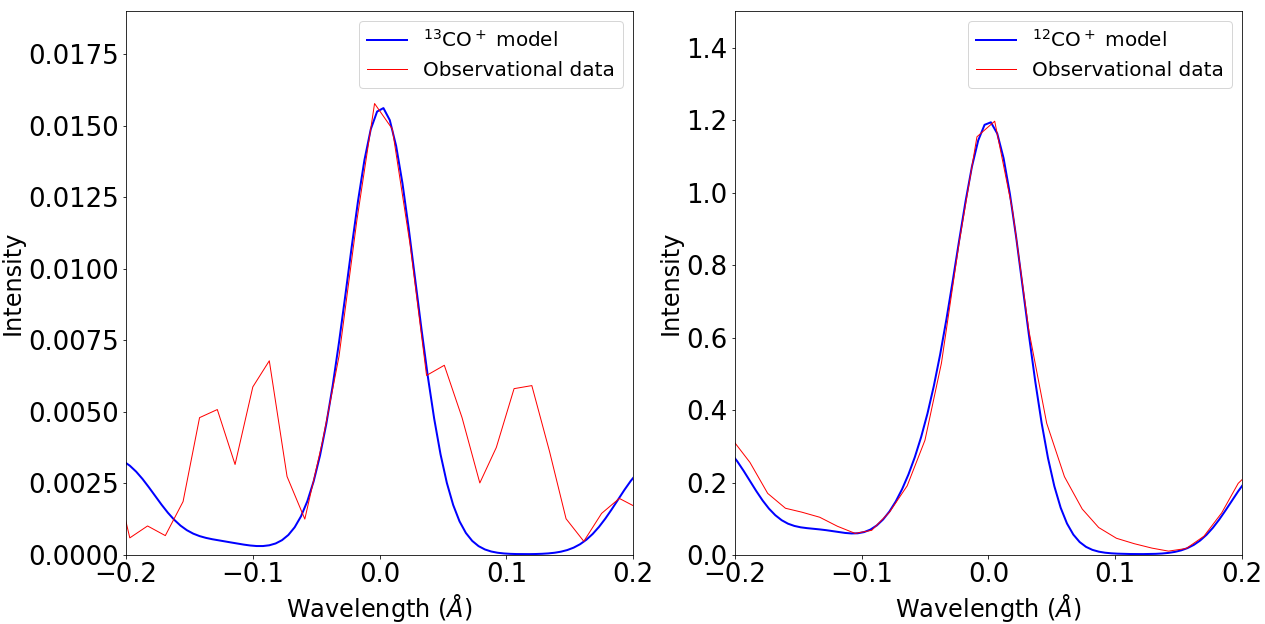}
\caption{Coaddition of the brightest (3,0) and (2,0) emission lines of $^{13}$CO$^+$ (left)
and $^{12}$CO$^+$ (right) with the corresponding modeling (see Table~\ref{tab:lines12C13Clist} for the
detailed list of coadded lines).}
\label{coaddlines}
\end{figure*}

Thanks to the coaddition of these lines we managed to detect a $^{13}$CO$^+$ coadded emission 
line well above the background noise. The coaddition of the $^{12}$CO$^+$ emission lines 
counterpart and their fit with our fluorescence model allowed us to derive the $^{12}$C/$^{13}$C 
isotopic ratio in the CO$^+$ by adjusting our $^{13}$CO$^+$ fluorescence
model (coaddition of the same lines) to the observational data (see Fig.~\ref{coaddlines}). 
This method provided $^{12}$C/$^{13}$C=$73\pm20$. The main source of uncertainty comes from 
the exact backround level that needs to be adjusted with a high accuracy. A good test for the 
adjustement of this level, apart that the intensities should be 
positive and a background close to zero in between the emission lines, is the quality of the fit with 
our modeling. The quality of this fit is very good for the FWHM if the model is adjusted to 
observational data with background  close to zero. If it is not the case, some discrepancies in the 
line width appear.

This estimate of the $^{12}$C/$^{13}$C isotopic ratio in CO$^+$ is similar to the other measurements
done in the optical in C$_2$ and CN \citep{manfroid:2009}, and in the submm region for
HCN. This ratio has been measured so far from ground-based facilities in dozens of comets
\citep{bockelee-morvan:2015}. Some other in situ measurements were more recently performed
in the coma of comet 67P/Churyumov-Gerasimenko by the ROSINA mass spectrometer on-board
the Rosetta spacecraft for C$_2$H$_4$, C$_2$H$_5$, CO \citep{rubin:2017}
and CO$_2$ \citep{hassig:2017}  molecules. These measurements varies from about 60 (measured in 
C$_2$ for comet 
West 1976 IV, \cite{lambert:1983}) to 165 (measured in CN with comet C/1995 O1 (Hale-Bopp) by
\cite{arpigny:2003}) with large errorbars. Most of them are compatible with the terrestrial value 
of 89, within
their (often large) errorbars. For CO in comet 67P, the $^{12}$C/$^{13}$C ratio measured is
86$\pm$9 \citep{rubin:2017}, i.e. is compatible with our result within the errorbars. 
This one is the first ever
published for the CO$^+$ species.

\section{Conclusion}

Thanks to high quality observational data obtained on comet C/2016 R2, a comet exceptionally
rich in both CO$^+$ and N$_2^+$ in the inner coma, it was possible to test in detail a new
fluorescence model of $^{12}$CO$^+$, that takes into account new laboratory data and theoretical
works obtained 
since the pioneering work of \cite{magnani:1986}. The confrontation of this model with our
spectra provides good results, confirming the efficiency of this model. Some new
fluorescence efficiencies (g-factors) are also published, which are in relatively 
good agreement with observational data for the relative intensities. Such revised g-factors 
can be used in the future
for more accurate measurement of CO$^+$ abundance in comets. Previous estimate of such abundances
can now also be slightly revised. For example the N$_2$/CO ratio computed by \cite{opitom:2019}
and based on the g-factor of the (2,0) band published by \cite{magnani:1986} can be 
revised (in addition to the revised g-factors published for N$_2^+$ by
\cite{rousselot:2022} and used by \cite{anderson:2022}). The new ratio CO$^+$/N$_2^+$ should
be reduced by about 20\% according to our revised values.

Our $^{13}$CO$^+$ fluorescence model also opens the possibility of new measurements of the 
$^{12}$C/$^{13}$C isotopic ratio in the CO$^+$ species. This work shows that such a measurement
becomes possible and provided a value of 73$\pm$20 for comet C/2016~R2, in agreement
with other similar measurements conducted in other species and in the in situ measurement
obtained in the coma of comet 67P by ROSINA mass spectrometer. This value probably excludes that 
the very peculiar comet C/2016 R2 would be of interstellar origin.

\begin{acknowledgements}
      Based on observations made with ESO Telescopes at the La
Silla Paranal Observatory under program 2100.C-5035(A).
E. Jehin is a Belgian FNRS Senior Research Associate, D. Hutsemékers is a 
Research Director, and J. Manfroid is a Honorary Research Director of the FNRS.
\end{acknowledgements}


\bibliographystyle{aa}
\bibliography{COplus}

\begin{thebibliography}{32}
\expandafter\ifx\csname natexlab\endcsname\relax\def\natexlab#1{#1}\fi

\bibitem[{{Anderson} {et~al.}(2022){Anderson}, {Rousselot}, {Noyelles},
  {Opitom}, {Jehin}, {Hutsem{\'e}kers}, \& {Manfroid}}]{anderson:2022}
{Anderson}, S.~E., {Rousselot}, P., {Noyelles}, B., {et~al.} 2022, \mnras, 515,
  5869

\bibitem[{{Arpigny}(1964{\natexlab{a}})}]{arpigny:1964b}
{Arpigny}, C. 1964{\natexlab{a}}, Annales d'Astrophysique, 27, 393

\bibitem[{{Arpigny}(1964{\natexlab{b}})}]{arpigny:1964a}
{Arpigny}, C. 1964{\natexlab{b}}, Annales d'Astrophysique, 27, 406

\bibitem[{{Arpigny} {et~al.}(2003){Arpigny}, {Jehin}, {Manfroid},
  {Hutsem{\'e}kers}, {Schulz}, {St{\"u}we}, {Zucconi}, \&
  {Ilyin}}]{arpigny:2003}
{Arpigny}, C., {Jehin}, E., {Manfroid}, J., {et~al.} 2003, Science, 301, 1522

\bibitem[{{Bell} {et~al.}(2007){Bell}, {Whyatt}, {Viti}, \&
  {Redman}}]{bell:2007}
{Bell}, T.~A., {Whyatt}, W., {Viti}, S., \& {Redman}, M.~P. 2007, \mnras, 382,
  1139

\bibitem[{{Billoux} {et~al.}(2014){Billoux}, {Cressault}, \&
  {Gleizes}}]{billoux:2014}
{Billoux}, T., {Cressault}, Y., \& {Gleizes}, A. 2014, \jqsrt, 133, 434

\bibitem[{{Biver} {et~al.}(2018){Biver}, {Bockel{\'e}e-Morvan}, {Paubert},
  {Moreno}, {Crovisier}, {Boissier}, {Bertrand}, {Boussier}, {Kugel}, {McKay},
  {Russo}, \& {DiSanti}}]{Biver:2018}
{Biver}, N., {Bockel{\'e}e-Morvan}, D., {Paubert}, G., {et~al.} 2018, \aap,
  619, A127

\bibitem[{{Bockel{\'e}e-Morvan} {et~al.}(2015){Bockel{\'e}e-Morvan},
  {Calmonte}, {Charnley}, {Duprat}, {Engrand}, {Gicquel}, {H{\"a}ssig},
  {Jehin}, {Kawakita}, {Marty}, {Milam}, {Morse}, {Rousselot}, {Sheridan}, \&
  {Wirstr{\"o}m}}]{bockelee-morvan:2015}
{Bockel{\'e}e-Morvan}, D., {Calmonte}, U., {Charnley}, S., {et~al.} 2015, \ssr,
  197, 47

\bibitem[{{Cochran} \& {McKay}(2018)}]{cochran:2018a}
{Cochran}, A.~L. \& {McKay}, A.~J. 2018, \apjl, 854, L10

\bibitem[{{Coxon} {et~al.}(2010){Coxon}, {K{\k{e}}pa}, \&
  {Piotrowska}}]{coxon:2010}
{Coxon}, J.~A., {K{\k{e}}pa}, R., \& {Piotrowska}, I. 2010, Journal of
  Molecular Spectroscopy, 262, 107

\bibitem[{{Ferchichi} {et~al.}(2022){Ferchichi}, {Derbel}, {Alijah}, \&
  {Rousselot}}]{ferchichi:2022}
{Ferchichi}, O., {Derbel}, N., {Alijah}, A., \& {Rousselot}, P. 2022, \aap,
  661, A132

\bibitem[{{Fowler}(1909{\natexlab{a}})}]{fowler:1909a}
{Fowler}, A. 1909{\natexlab{a}}, \mnras, 70, 176

\bibitem[{{Fowler}(1909{\natexlab{b}})}]{fowler:1909b}
{Fowler}, A. 1909{\natexlab{b}}, \mnras, 70, 179

\bibitem[{{Fowler}(1910)}]{fowler:1910}
{Fowler}, A. 1910, \mnras, 70, 484

\bibitem[{{Hakalla} {et~al.}(2019){Hakalla}, {Szajna}, {Piotrowska}, {Malicka},
  {Zachwieja}, \& {K{\k{e}}pa}}]{hakalla:2019}
{Hakalla}, R., {Szajna}, W., {Piotrowska}, I., {et~al.} 2019, \jqsrt, 234, 159

\bibitem[{{H{\"a}ssig} {et~al.}(2017){H{\"a}ssig}, {Altwegg}, {Balsiger},
  {Berthelier}, {Bieler}, {Calmonte}, {Dhooghe}, {Fiethe}, {Fuselier}, {Gasc},
  {Gombosi}, {Le Roy}, {Luspay-Kuti}, {Mandt}, {Rubin}, {Tzou}, {Wampfler}, \&
  {Wurz}}]{hassig:2017}
{H{\"a}ssig}, M., {Altwegg}, K., {Balsiger}, H., {et~al.} 2017, \aap, 605, A50

\bibitem[{{Herzberg, G.}(1950)}]{Herzberg:1950}
{Herzberg, G.} 1950, {Molecular Spectra and Molecular Structure. I. Spectra of
  Diatomic Molecules}, ed. V.~N.~R. company

\bibitem[{{Kepa} {et~al.}(2002){Kepa}, {Kocan}, {Ostrowska},
  {Piotrowska-Domagala}, {Jakubek}, \& {Zachwieja}}]{kepa:2002}
{Kepa}, R., {Kocan}, A., {Ostrowska}, M., {et~al.} 2002, Journal of Molecular
  Spectroscopy, 214, 117

\bibitem[{{K{\k{e}}pa} {et~al.}(2004){K{\k{e}}pa}, {Kocan},
  {Ostrowska-Kope{\'c}}, {Piotrowska-Domaga{\l}a}, \& {Zachwieja}}]{kepa:2004}
{K{\k{e}}pa}, R., {Kocan}, A., {Ostrowska-Kope{\'c}}, M.,
  {Piotrowska-Domaga{\l}a}, I., \& {Zachwieja}, M. 2004, Journal of Molecular
  Spectroscopy, 228, 66

\bibitem[{{Kurucz} {et~al.}(1984){Kurucz}, {Furenlid}, {Brault}, \&
  {Testerman}}]{kurucz:1984}
{Kurucz}, R.~L., {Furenlid}, I., {Brault}, J., \& {Testerman}, L. 1984, {Solar
  flux atlas from 296 to 1300 nm}

\bibitem[{{Lambert} \& {Danks}(1983)}]{lambert:1983}
{Lambert}, D.~L. \& {Danks}, A.~C. 1983, \apj, 268, 428

\bibitem[{{Magnani} \& {A'Hearn}(1986)}]{magnani:1986}
{Magnani}, L. \& {A'Hearn}, M.~F. 1986, \apj, 302, 477

\bibitem[{{Manfroid} {et~al.}(2009){Manfroid}, {Jehin}, {Hutsem{\'e}kers},
  {Cochran}, {Zucconi}, {Arpigny}, {Schulz}, {St{\"u}we}, \&
  {Ilyin}}]{manfroid:2009}
{Manfroid}, J., {Jehin}, E., {Hutsem{\'e}kers}, D., {et~al.} 2009, \aap, 503,
  613

\bibitem[{{Opitom} {et~al.}(2019){Opitom}, {Hutsem{\'e}kers}, {Jehin},
  {Rousselot}, {Pozuelos}, {Manfroid}, {Moulane}, {Gillon}, \&
  {Benkhaldoun}}]{opitom:2019}
{Opitom}, C., {Hutsem{\'e}kers}, D., {Jehin}, E., {et~al.} 2019, \aap, 624, A64

\bibitem[{{Rosmus} \& {Werner}(1982)}]{rosmus:1982}
{Rosmus}, P. \& {Werner}, H.-J. 1982, Molecular Physics, 47, 661

\bibitem[{{Rousselot} {et~al.}(2022){Rousselot}, {Anderson}, {Alijah},
  {Noyelles}, {Opitom}, {Jehin}, {Hutsem{\'e}kers}, \&
  {Manfroid}}]{rousselot:2022}
{Rousselot}, P., {Anderson}, S.~E., {Alijah}, A., {et~al.} 2022, \aap, 661,
  A131

\bibitem[{{Rubin} {et~al.}(2017){Rubin}, {Altwegg}, {Balsiger}, {Berthelier},
  {Bieler}, {Calmonte}, {Combi}, {De Keyser}, {Engrand}, {Fiethe}, {Fuselier},
  {Gasc}, {Gombosi}, {Hansen}, {H{\"a}ssig}, {Le Roy}, {Mezger}, {Tzou},
  {Wampfler}, \& {Wurz}}]{rubin:2017}
{Rubin}, M., {Altwegg}, K., {Balsiger}, H., {et~al.} 2017, \aap, 601, A123

\bibitem[{{Swings}(1965)}]{swings:1965}
{Swings}, P. 1965, \qjras, 6, 28

\bibitem[{{Szajna} {et~al.}(2004){Szajna}, {K{\k{e}}pa}, \&
  {Zachwieja}}]{Szajna:2004}
{Szajna}, W., {K{\k{e}}pa}, R., \& {Zachwieja}, M. 2004, European Physical
  Journal D, 30, 49

\bibitem[{{Weryk} \& {Wainscoat}(2016)}]{weryk:2016}
{Weryk}, R. \& {Wainscoat}, R. 2016, Central Bureau Electronic Telegrams, 4318

\bibitem[{{Wierzchos} \& {Womack}(2018)}]{Wierzchos:2018}
{Wierzchos}, K. \& {Womack}, M. 2018, ArXiv e-prints
  [\eprint[arXiv]{1805.06918}]

\bibitem[{{Zucconi} \& {Festou}(1985)}]{zucconi:1985}
{Zucconi}, J.~M. \& {Festou}, M.~C. 1985, \aap, 150, 180

\end{thebibliography}

\end{document}